\documentclass[pra,aps,twocolumn,showpacs]{revtex4}

\usepackage{epsfig}
\usepackage{amssymb}
\usepackage{graphicx}

\begin{document}

\sloppy

\title{Instantaneous Measurement of field quadrature moments and entanglement}

\author{P. Lougovski,$^1$ H. Walther,$^1$ and E. Solano$^{1,2}$}

\address{$^1$Max-Planck-Institut f{\"u}r Quantenoptik,
Hans-Kopfermann-Strasse 1, D-85748 Garching, Germany\\
$^2$Secci\'{o}n F\'{\i}sica, Departamento de Ciencias, Pontificia
Universidad Cat\'{o}lica del Per\'{u}, Apartado 1761, Lima, Peru }

\begin{abstract}
We present a method of measuring expectation values of quadrature
moments of a multimode field through two-level probe
``homodyning''. Our approach is based on an integral transform
formalism of measurable probe observables, where analytically
derived kernels unravel efficiently the required field information
at zero interaction time, minimizing decoherence effects. The
proposed scheme is suitable for fields that, while inaccessible to
a direct measurement, enjoy one and two-photon Jaynes-Cummings
interactions with a two-level probe, like spin, phonon, or cavity
fields. Available data from previous experiments are used to
confirm our predictions.
\end{abstract}


\maketitle

State reconstruction of a bosonic field is an important issue in
fundamentals of quantum physics that has been studied extensively,
both theoretically and experimentally, in the last two
decades~\cite{SchleichBook}. Its main concern is to measure either
the density matrix of an unknown quantum field state or,
equivalently, any of its phase-space representations. Among them,
Wigner function reconstructions seem to be the most promising
avenue for measuring completely, for example, an intracavity
microwave field~\cite{Bertet} or the motion of a trapped
ion~\cite{Leibfried}. In the case of a propagating field, usually
accessible to direct measurement, homodyne techniques are
currently used in the lab~\cite{MandelWolf}. Typically, a complete
state reconstruction with standard techniques demand great
experimental efforts, is strongly affected by decoherence
mechanisms, and, frequently, the obtained information exceeds our
requirements. In those cases, techniques for extracting
efficiently the required partial information are most welcomed and
even necessary. The problem is even harder when the field is not
directly accessible and a quantum probe has to be used for the
purposes of an indirect measurement~\cite{Brune}. Therefore, the
following question arises: how to derive accurately the
expectation value of a field observable, through an efficient
measure of a probe, with minimal resources and with an outcome
that is minimally affected by decoherence mechanisms? In this
article, we answer this question for the case of a multimode
bosonic field, interacting with a two-level probe, by means of a
practical integral transform method. These conditions are
naturally fulfilled by several physical systems, like a cavity
field interacting with two-level atoms, the motion of an ion
interacting, through laser coupling, with two of its internal
levels, or even several spins, in a mode approximation,
interacting with a single spin, like in NMR or quantum dot
systems.

We consider a general picture in which an inaccessible bosonic
field is measured through an interacting probe, following the
interaction Hamiltonian
\begin{eqnarray}
\label{hamiltonian} H = \hbar \sum_{i,j} g_{i,j} ( p_i
f_j^{\dagger} + p_i^{\dagger} f_j ) ,
\end{eqnarray}
where $p_i$ and $f_j$ are probe and field operators, respectively,
and $g_{i,j}$ are coupling strengths. We postulate the existence
of an analytical kernel $\kappa (\tau)$ such that
\begin{eqnarray}
\label{integraltransform} \langle F \rangle =
\int_{-\infty}^{\infty} \kappa (\tau) \langle P \rangle (\tau) d
\tau ,
\end{eqnarray}
where $F$ and $P$ are operators associated with field and probe
observables, respectively, and $\tau$ is the dimensionless
probe-field interaction time. Later, it will be clear why it is
possible to include a negative axis for $\tau$ in the integration
limits of Eq.~(\ref{integraltransform}). $\langle P \rangle
(\tau)$ will be replaced by experimental measured data and, for
the method to be useful, we should be able to formally invert the
integral transform and derive an analytical expression for
$\kappa(\tau)$. We will show below that this inversion is possible
for several important field observables, like quadrature moments
of a multimode field, unravelling important information on
squeezing and entanglement. Note that we aim at measuring
efficiently partial field information without the requirement of
full state reconstruction, even though integral techniques can
also provide us with complete Wigner function
reconstructions~\cite{Fresnel}.

We consider a two-level probe interacting with a single-mode field
through a resonant Jaynes-Cummings (JC) Hamiltonian, in the
interaction picture,
\begin{eqnarray}
\label{jaynescummings} H_{\rm JC} = \hbar g ( \sigma^{\dagger} a +
\sigma a^{\dagger} ) ,
\end{eqnarray}
where $g$ is a coupling strength, $\{ \sigma , \sigma^{\dagger}
\}$ are lowering and raising probe operators, and $\{ a ,
a^{\dagger} \}$ are annihilation and creation field operators. We
assume that, given that the initial probe-field density operator
is $\rho^{\rm in} (0)$, we can measure,  after a dimensionless
interaction time $\tau \equiv g t$, the population of the excited
probe level $| e \rangle$
\begin{eqnarray}
\label{expectationobservable}  P_{\rm e}^{\rm in} (\tau) \equiv
{\rm Tr} \lbrack \rho (\tau) | e \rangle \langle e | \rbrack =
{\rm Tr} \lbrack U (\tau) \rho^{\rm in} (0) U^{\dagger} (\tau) | e
\rangle \langle e | \rbrack ,
\end{eqnarray}
where  $| e \rangle \langle e | = \sigma^{\dagger} \sigma$ and $U
(\tau) = \exp ( -i \tau H_{\rm JC} / \hbar g )$ is the evolution
operator. In Eq.~(\ref{expectationobservable}), and throughout
this work, upper and lower indices stem from initial and measured
probe states, respectively. We consider the initial state
$\rho^{+_{\phi}} = | +_{\phi} \rangle \langle +_{\phi} | \otimes
\rho_{f}$, where $| \pm_{\phi} \rangle = ( | g \rangle \pm e^{i
\phi} | e \rangle ) / \sqrt{2}$ are the eigenvectors of
$\sigma^{\phi}_x = \sigma^{\dagger} e^{i \phi} + \sigma e^{-i
\phi}$ with $\sigma^{\phi}_x | \pm_{\phi} \rangle = \pm |
\pm_{\phi} \rangle$. When $\phi = 0$, $\sigma^{\phi}_x$ turns into
$\sigma_x$, the usual spin-$1/2$ Pauli operator. Replacing
$\rho^{+_{\phi}}$ in Eq.~(\ref{expectationobservable}) gives
\begin{eqnarray}
\label{probabilityprobe} && \!\!\!\!\!\!\!\! P_{\rm e}^{+_{\phi}}
(\tau) = \! \frac{i}{4} \! \sum^{\infty}_{n=0} \sin(2 \tau \sqrt{n
+ 1}) (e^{i\phi}\rho_{n , n+1} - e^{-i\phi} \rho_{n+1 , n})
\nonumber
\\ && + \frac{1}{2} \sum^{\infty}_{n=0}
(\cos^{2}(\tau \sqrt{n+1})\rho_{n,
 n} +  \sin^{2}(\tau \sqrt{n+1})\rho_{n+1 , n+1}) , \nonumber
\\ &&
\end{eqnarray}
where the initial field state $\rho_f = \sum_{n,m} \rho_{n,m} | n
\rangle \langle m |$ has been written in terms of its matrix
elements. Through the knowledge of $P_{\rm e}^{+_{\phi}} (\tau)$,
we aim at measuring the field quadratures $X_{\phi} = \frac{1}{2}
( a e^{-i \phi} + a^{\dagger} e^{i \phi} )$ and $Y_{\phi} =
X_{\phi + \pi / 2} = \frac{1}{2i} ( a e^{-i \phi} - a^{\dagger}
e^{i \phi} )$ with expectation values
\begin{eqnarray}
\label{quadraturex} && \!\!\!\!\!\!\!\!\!\!\!\! \langle X_{\phi}
\rangle = \frac{1}{2} \sum^{\infty}_{n=0} \sqrt{n+1}(e^{i \phi
}\rho_{n , n+1} + e^{-i \phi}\rho_{n+1 , n}) \\
\label{quadraturey} && \!\!\!\!\!\!\!\!\!\!\! \langle Y_{\phi}
\rangle = \frac{i}{2} \sum^{\infty}_{n=0} \sqrt{n+1} (e^{i
\phi}\rho_{n , n+1} - e^{- i \phi}\rho_{n+1 , n}) .
\end{eqnarray}
We choose the kernel in Eq.~(\ref{integraltransform}) as an odd
function, $\kappa ( -\tau ) = - \kappa (\tau)$, so that the
integral of the second sum in Eq.~(\ref{probabilityprobe})
vanishes, while the integral of the first sum should reproduce
Eq.~(\ref{quadraturey}). In consequence, replacing operators $P
\rightarrow | e \rangle \langle e |$ and $F \rightarrow Y_{\phi}$
in Eq.~(\ref{integraltransform}), the condition for this ansatz to
be true is
\begin{eqnarray}
\label{ansatz} \int_{-\infty}^{\infty} \kappa (\tau) e^{i \omega_n
\tau} d \tau = i \omega_n ,
\end{eqnarray}
where $\sin(2 \tau \sqrt{n + 1})$ has been rewritten in complex
form with $w_n = 2 \sqrt{n + 1}$. The inverse Fourier transform of
Eq.~(\ref{ansatz}) provides us with the kernel
\begin{eqnarray}
\label{kernel} \kappa (\tau) = \frac{i}{2 \pi}
\int_{-\infty}^{\infty} e^{-i \omega_n \tau} \omega_n d \omega_n =
- \delta'(\tau) ,
\end{eqnarray}
where $\delta'(\tau)$ is the first derivative of a delta function.
Note that even if, physically, $w_n$ is a function of discrete
$n$'s, it can be treated formally as continuous for the sake of
the inverse transform. Then, Eq.~(\ref{integraltransform}) can be
written as
\begin{eqnarray}
\langle Y_{\phi} \rangle = - \int_{-\infty}^{\infty} \delta'
(\tau) P_{\rm e}^{+_{\phi}} (\tau) d \tau ,
\end{eqnarray}
yielding
\begin{eqnarray}
\label{firstresulty} \langle Y_{\phi} \rangle = \frac{d}{d \tau}
P_{\rm e}^{+_{\phi}} (\tau) \bigg\vert_{\tau = 0} ,
\end{eqnarray}
where the continuity of the first derivative of $P_{\rm
e}^{+_{\phi}} (\tau)$ at $\tau = 0$ has been considered.
Similarly,
\begin{eqnarray}
\label{firstresultx} \langle X_{\phi} \rangle = \langle Y_{\phi -
\frac{\pi}{2}} \rangle =\frac{d}{d \tau} P_{\rm e}^{+_{\phi -
\frac{\pi}{2}}} (\tau) \bigg\vert_{\tau = 0} .
\end{eqnarray}

Eqs.~(\ref{firstresulty}) and (\ref{firstresultx}) show that
$\langle X_{\phi} \rangle$ and $\langle Y_{\phi} \rangle$ are
fully contained in the first derivative, at $\tau = 0$, of the
measured probe population, offering a remarkably simple way of
obtaining quadrature information. Note that knowing the first
derivative at $\tau = 0$ requires knowing the function in a
vicinity. However, no necessity of full state reconstruction or
lengthy time integration over Rabi oscillations are needed, in
contrast to standard methods. Needless to say, the influence of
decoherence is minimized.

Induced by the structure of Eq.~(\ref{probabilityprobe}), and
aiming at cancelling the population while keeping the off-diagonal
(phase) information, we could find a similar result by subtracting
rotated populations
\begin{eqnarray}
\label{probabilityhomodyning} && \!\!\!\!\!\!\!\!\!\!\!\!\!\!\!
P_{\rm e}^{+_{\phi}} (\tau) - P_{\rm e}^{-_{\phi}} (\tau)  =
\nonumber
\\ && \!\!\!\!\! \frac{i}{4} \sum^{\infty}_{n=0} \sin(2 \tau
\sqrt{n + 1}) (e^{i\phi}\rho_{n , n+1} - e^{-i\phi} \rho_{n+1 ,
n}) .
\end{eqnarray}
Following a similar procedure as before, we can write
\begin{eqnarray}
\label{homodyningy} \langle Y_{\phi} \rangle = \frac{1}{2} \bigg(
\frac{d}{d \tau} P_{\rm e}^{+_{\phi}} (\tau) - \frac{d}{d \tau}
P_{\rm e}^{-_{\phi}} (\tau) \bigg) \bigg\vert_{\tau = 0} .
\end{eqnarray}
This result has evident resemblance to the known technique of
field homodyning~\cite{MandelWolf}. There, an unknown field is
mixed in a 50-50 beam splitter with a local oscillator, and the
difference of field intensities (rate of photon clicks) at the
output gives us quadrature information. Based on this similarity,
the proposed method could be called after two-level probe
"homodyning".

$X_{\phi}$ and $Y_{\phi}$ happen to be relevant observables in a
wide range of physical systems where current experiments enjoy
probe rotations and JC-like interactions, like cavity QED (CQED),
trapped ions, BEC, and different solid-state systems. In CQED, the
quadrature information can be obtained by sending an excited atom
through a Ramsey zone before crossing the cavity
mode~\cite{ParisReview}, and finally measuring the population of
the excited state at the cavity output. For trapped ions, $\langle
X \rangle$ and $\langle Y \rangle$ represent, literally,
expectation values of position and momentum operators, that will
be obtained by measuring the internal state statistics, where the
efficiency can reach $\sim 100 \%$, after a JC-like sideband
excitation~\cite{IonsReview}. In the case of solid-state devices,
there are several systems enjoying two-level probes interacting
through JC interactions with cavity, phonon or spin fields. It is
noteworthy to mention that in all these examples a probe is needed
due to the lack of a direct measurement.

In the rest of this article, for the sake of simplicity, we will
use the language of cavity QED, where a two-level atom probes an
intracavity electromagnetic field.

Another important field observable that can be obtained
straightforwardly with a JC interaction is the mean photon number
$\langle n \rangle = \langle a^{\dagger} a \rangle$. Considering
the initial state $\rho^{\rm e} = | e \rangle \langle e | \otimes
\rho_f$, we can derive the kernel
\begin{eqnarray}
{\bar \kappa} (\tau) = - \delta''(\tau)
\end{eqnarray}
for measuring
\begin{equation}\label{meann}
\langle n \rangle = \frac{1}{2} \frac{d^{2}P_{\rm g}^{\rm e}
(\tau)}{d^{2}\tau} \Big|_{\tau=0} - 1 \,\, .
\end{equation}
Note that measuring $\langle n \rangle$ does not require Ramsey
zones for rotating the atom, as was the case before. Given the
available experimental data, expression in Eq.~(\ref{meann}) is
the only one that could be presently tested. For example, using
the experimental data associated with the experiments at ENS, see
Figs.~2 (A), (B) in Ref.~\cite{Brune}, we predict $\langle n
\rangle \approx 0.14$ and $0.81$, respectively. These values are
quite close to the ones obtained via integration or fitting long
Rabi oscillations, 0.06 and 0.85, respectively. We made similar
estimations for the experiments at NIST, obtaining $\langle n
\rangle \approx 1.6$ and $3.1$ for the experiments associated with
Figs.~2 and 3 in Ref.~\cite{Wineland}, to be compared with 1.5 and
2.9, respectively. Clearly, our predictions could only be better
if specific experiments are performed, aiming at first and second
derivatives at very short interaction times.

It is also possible to use these integral methods to measure
second-order quadrature moments, providing information about field
quadrature squeezing and entanglement of a multimode field. We
will use a resonant two-photon JC Hamiltonian that reads
\begin{eqnarray}
\label{twophotonjc} H_{\rm 2JC} = \hbar g ( \sigma^{\dagger} a^2 +
\sigma a^{\dagger \, 2} )
\end{eqnarray}
in the interaction picture. This nonlinear interaction has been
realized experimentally in the context of microwave
CQED~\cite{Haroche} and trapped ions~\cite{Wineland}. Our aim,
here, is to measure expectation values of squared quadratures,
\begin{eqnarray}
\label{quadraturex2} \langle X^{2}_{\phi} \rangle && \!\!\!\!\!\!
= \frac{1}{4} + \frac{ \langle n \rangle }{2} \nonumber \\
&& \!\!\!\!\!\!\!\!\!\!\!\! + \frac{1}{4} \sum^{\infty}_{n=0}
\sqrt{(n+1)(n+2)} \big( e^{2i\phi}\rho_{n , n+2} +  e^{-2i\phi}
\rho_{n+2 , n} \big)
, \nonumber \\ \\
\label{quadraturey2} \langle Y^{2}_{\phi} \rangle && \!\!\!\!\!\!
= \frac{1}{4} + \frac{ \langle n \rangle }{2} \nonumber \\
&& \!\!\!\!\!\!\!\!\!\!\!\! - \frac{1}{4} \sum^{\infty}_{n=0}
\sqrt{(n+1)(n+2)} \big( e^{2i\phi}\rho_{n , n+2} +  e^{-2i\phi}
\rho_{n+2 , n} \big)
  . \nonumber \\
\end{eqnarray}
with the help of Eq.~(\ref{twophotonjc}) and the proposed integral
transform techniques. Then, in a close analogy to
Eq.~(\ref{probabilityhomodyning}), now for a two-photon JC
interaction, we can calculate
\begin{eqnarray}
\label{prodif2ph} \!\!\!\!\!\! P^{+_{\phi}}_{\rm g}(\tau) \! - \!
P^{-_{\phi}}_{\rm g}(\tau) & \!\! = \!\! & \frac{i}{4}
\sum^{\infty}_{n=0} \sin(2 \tau \sqrt{(n+1)(n+2)}) \nonumber\\ &
& \times (e^{2i\phi}\rho_{n , n+2} + e^{-2i\phi}\rho_{n+2 , n}).
\end{eqnarray}
With the help of Eqs.~(\ref{meann}), (\ref{quadraturex2}), and
(\ref{quadraturey2}), and by deriving and using the corresponding
kernels, we arrive at
\begin{eqnarray}
\label{squareX} \langle X^{2}_{\phi} \rangle & = & \frac{1}{2i}
\big( \frac{d P^{+_{\phi}}_{\rm g}(\tau)}{d\tau} -
\frac{d P^{-_{\phi}}_{\rm g}(\tau)}{d \tau} \big) \Big|_{\tau=0}
\nonumber\\
& & + \frac{1}{4} \frac{d^{2}P_{\rm g}^{\rm e} (\tau)}{d^{2}\tau}
\Big|_{\tau=0} - \frac{1}{4} , \\
\label{squareY} \langle Y^{2}_{\phi} \rangle & = & \frac{i}{2}
\big( \frac{d P^{+_{\phi}}_{\rm g}(\tau)}{d\tau} -
\frac{d P^{-_{\phi}}_{\rm g} (\tau)}{d \tau} \big) \Big|_{\tau=0}
\nonumber\\
& & + \frac{1}{4} \frac{d^{2}P_{\rm g}^{\rm e} (\tau)}{d^{2}\tau}
\Big|_{\tau=0} - \frac{1}{4} .
\end{eqnarray}
The quadrature variances $(\Delta X)^{2}= \langle X^{2} \rangle -
\langle X \rangle^{2}$ and $(\Delta Y)^{2}= \langle Y^{2} \rangle
- \langle Y \rangle^{2}$ contain information about field squeezing
and can be calculated straightforwardly by using
Eqs.~(\ref{firstresulty}), (\ref{firstresultx}), (\ref{squareX}),
and (\ref{squareY}).

It is noteworthy to say that it is not necessary to use a
two-photon JC interaction for measuring second-order quadrature
moments. For example, it is enough to use a two-atom probe
interacting with the tested field through a single-photon JC,
whose interaction Hamiltonian reads
\begin{eqnarray}
\label{twoatomjc} H^{\rm I} = \hbar g \lbrack ( \sigma^{\dagger}_1
+ \sigma^{\dagger}_2 )  a + ( \sigma_1 + \sigma_2 ) a^{\dagger}
\rbrack ,
\end{eqnarray}
where the subindexes are labelling probe atoms ``$1$'' and
``$2$''. We consider the Bell states $| \phi^{+}_{\theta} \rangle
= \lbrack | g_1 g_2 \rangle + e^{i \theta} | e_1 e_2 \rangle
\rbrack / \sqrt{2}$ and $| \phi^{-}_{\theta} \rangle = \lbrack |
g_1 g_2 \rangle - e^{i \theta} | e_1 e_2 \rangle \rbrack /
\sqrt{2}$ as two probe initial states, and in both cases we
measure $| \psi^{+} \rangle = \lbrack | g_1 e_2 \rangle + | e_1
g_2 \rangle \rbrack / \sqrt{2}$, obtaining
\begin{eqnarray}
\label{twoatomhomodyning} P^{\phi^{+}_{\theta} }_{ \psi^{+} }
(\tau) - P^{ \phi^{-}_{\theta} }_{ \psi^{+} } (\tau) =
\frac{1}{2} \sum^{\infty}_{n=0} \frac{\sqrt{(n+1)(n+2)}}{2n+3}
\nonumber \\
\times \sin^{2}(\sqrt{2} \tau \sqrt{2n + 3}) (e^{i\theta}\rho_{n ,
n+2} + e^{-i\theta} \rho_{n+2 , n}) .
\end{eqnarray}
From this expression, and following similar steps to previous
derivations, it is possible to deduce
\begin{eqnarray}\label{twoatomsquarex}
\langle X^{2}_{\theta} \rangle = \frac{1}{8} \bigg(
\frac{d^{2}P^{\phi^{+}_{\theta} }_{ \psi^{+} } (\tau)}{d^{2}\tau}
- \frac{d^{2}P^{\phi^{-}_{\theta} }_{ \psi^{+} }
(\tau)}{d^{2}\tau} \bigg) \Big|_{\tau=0} \nonumber \\
+ \frac{1}{4} \frac{d^{2}P_{g}(\tau)}{d^{2}\tau} \Big|_{\tau=0} -
\frac{1}{4} , \\ \label{twoatomsquarey} \langle Y^{2}_{\theta}
\rangle = - \frac{1}{8} \bigg( \frac{d^{2}P^{\phi^{+}_{\theta} }_{
\psi^{+} } (\tau)}{d^{2}\tau} - \frac{d^{2}P^{\phi^{-}_{\theta}
}_{ \psi^{+} }
(\tau)}{d^{2}\tau} \bigg) \Big|_{\tau=0} \nonumber \\
+ \frac{1}{4} \frac{d^{2}P_{g}(\tau)}{d^{2}\tau} \Big|_{\tau=0} -
\frac{1}{4} \, \, .
\end{eqnarray}
Note that the required Bell states and the measurement procedure
have already been implemented in the lab in the case of
CQED~\cite{HarocheZheng} and trapped ion~\cite{WinelandMolmer}
setups.

The formalism for measuring squeezing can be generalized to a
two-mode field (or more), providing us with entanglement
information. Accordingly, we define the two-mode quadratures as
\begin{eqnarray}
X_{\phi} & = & X_{\phi_1} + X_{\phi_2} =
\frac{1}{2}\sum^{2}_{j=1}(a_{j}^{\dagger}e^{-i \phi_j} +
a_{j}e^{i \phi_j}) , \\
Y_{\phi} & = & Y_{\phi_1} + Y_{\phi_2} =
\frac{i}{2}\sum^{2}_{j=1}(a_{j}^{\dagger}e^{-i\phi_j} -
a_{j}e^{i\phi_j}),
\end{eqnarray}
where $j$ labels modes "1" and "2". The quantities $\langle
X_{\phi} \rangle$ and $\langle Y_{\phi} \rangle$ can be easily
calculated and, here, we will concentrate on the second-order
quadrature moments
\begin{eqnarray}
\langle X_{\phi}^{2} \rangle & = & \langle X_{\phi_1}^{2} \rangle
+ \langle X_{\phi_2}^{2}
\rangle + 2 \langle X_{\phi_1}X_{\phi_2} \rangle  \\
\langle Y_{\phi}^{2} \rangle & = & \langle Y_{\phi_1}^{2} \rangle
+ \langle Y_{\phi_2}^{2} \rangle + 2 \langle Y_{\phi_1} Y_{\phi_2}
\rangle,
\end{eqnarray}
In these expressions, we define $\langle X_{\phi_1} X_{\phi_2}
\rangle = \frac{1}{2} \langle A \rangle + \frac{1}{2} \langle B
\rangle$, $\langle Y_{\phi_1} Y_{\phi_2} \rangle =  \frac{1}{2}
\langle A \rangle - \frac{1}{2} \langle B \rangle$, with
\begin{eqnarray}
\label{quantA}A & = &
a_{1}^{\dagger}a_{2}e^{-i(\phi_{1}-\phi_{2})} +
a_{1}a_{2}^{\dagger}e^{i(\phi_{1}-\phi_{2})} , \\
\label{quantB}B & = &
a_{1}^{\dagger}a_{2}^{\dagger}e^{-i(\phi_{1}+\phi_{2})} +
a_{1}a_{2}e^{i(\phi_{1}+\phi_{2})} .
\end{eqnarray}
Single-mode quantities $\langle X_{\phi_i} \rangle$, $\langle
Y_{\phi_i} \rangle$, $\langle X_{\phi_i}^{2} \rangle$ and $\langle
Y_{\phi_i}^{2} \rangle$, can be determined using two-level probes
as it was shown above. Therefore, the main issue is to calculate
the expectation values of $A$ and $B$, which describe correlations
between modes $1$ and $2$. It has been shown,
theoretically~\cite{Neto} and experimentally~\cite{Haroche}, that
the two-photon probe-field interaction Hamiltonian
\begin{equation}\label{interHA}
H_{\rm A} = \hbar g(\sigma^{\dagger}a_{1}a_{2}^{\dagger} + \sigma
a_{1}^{\dagger}a_{2})
\end{equation}
can be engineered and controlled. If the probe is prepared
initially in the superposition states, $| +_{\phi} \rangle$ or $|
-_{\phi} \rangle$, with $\phi=\phi_{1}-\phi_{2}$, we can calculate
\begin{eqnarray}\label{prodif2m}
P^{+}_{\rm e,A}(\tau) - P^{-}_{\rm e,A}(\tau) & = &
\frac{i}{2}\sum^{\infty}_{n_{1},n_{2}=0}\sin(2g\tau\sqrt{n_{2}
(n_{1}+1)}) \nonumber\\ &  & \hspace*{-2.8cm} \times
(e^{-i\phi}\rho_{n_{1},n_{2};n_{1}+1,n_{2}-1} +
e^{i\phi}\rho_{n_{1}+1,n_{2}-1;n_{1},n_{2}}) ,
\end{eqnarray}
from which we can derive
\begin{eqnarray}
\label{Amean} \langle A \rangle & = &
\frac{i}{g}\left(\frac{dP^{+}_{\rm e,A}(\tau)}{d\tau} -
\frac{P^{-}_{\rm e,A}(\tau)}{d\tau}\right) \Big|_{\tau=0}.
\end{eqnarray}
Similarly, and by using the Hamiltonian
\begin{equation}\label{interHB}
H_{B} = \hbar g(\sigma^{\dagger}a_{1}^{\dagger}a_{2}^{\dagger} +
\sigma a_{1}a_{2}),
\end{equation}
we can deduce
\begin{eqnarray}\label{Bmean} \langle B \rangle & = &
\frac{i}{g}\left(\frac{dP^{+}_{\rm e,B}(\tau)}{d\tau} -
\frac{P^{-}_{\rm e,B}(\tau)}{d\tau}\right) \Big|_{\tau=0}.
\end{eqnarray}
In consequence, we can also estimate two-mode field variances
$(\Delta X)^{2}$ and $(\Delta Y)^{2}$ in terms of measurable probe
observables. Furthermore, by using the same approach, we can
compute the variances of EPR-like operators
\begin{eqnarray}
u & = & a_{0}X_{1} - \frac{c_{1}}{|c_{1}|}\frac{1}{a_{0}}X_{2} , \\
v & = & a_{0}Y_{1} - \frac{c_{2}}{|c_{2}|}\frac{1}{a_{0}}Y_{2},
\end{eqnarray}
where $a_{0}$, $c_{1}$ and $c_{2}$ are constants. For example, it
was shown in Ref.~\cite{Duan} that a two-mode Gaussian state
$\rho$ is separable if, and only if, $\langle (\Delta u)^{2}
\rangle_{\rho} + \langle (\Delta v)^{2} \rangle_{\rho} \ge a_{0}^2
+ 1 / a_{0}^2$.

In summary, we have shown how expectation values of quadrature
field operators can be measured by means of a two-level probe,
helped by a practical integral transform method and without the
necessity of full state reconstruction. Surprisingly, all relevant
information is contained in first and second derivatives of
measurable probe observables at interaction time $\tau = 0$,
making unnecessary long range probe measurements and minimizing
decoherence effects. Also, we showed that a similar technique
allows to measure second-order quadrature moments and variances,
that is, squeezing and entanglement. These results allow us to
conjecture the possibility of realizing full state reconstructing
with ``instantaneous'' measurements that are robust to
decoherence.

We thank C. Monroe, D. Wineland, J.-M. Raimond and S. Haroche for
providing us with useful experimental data. P. L. acknowledges
financial support from the Bayerisches Staatsministerium f\"ur
Wissenschaft, Forschung und Kunst in the frame of the Information
Highway Project and E. S. from the EU through RESQ project.

\end{document}